\documentclass[journal,comsoc,12pt]{IEEEtran}
\usepackage{amsmath,amsfonts}
\usepackage{algorithmic}
\usepackage{algorithm}
\usepackage{array}
\usepackage{textcomp}
\usepackage{stfloats}
\usepackage{url}
\usepackage{verbatim}
\usepackage{graphicx}
\usepackage{cite}
\usepackage[rgb]{xcolor}
\begin{document}

\title{Rate-Splitting Multiple Access and its Interplay with Intelligent Reflecting Surfaces}


\author{Arthur S. de Sena, \textit{Student Member}, \textit{IEEE}, Pedro H. J. Nardelli,  \textit{Senior Member}, \textit{IEEE}, Daniel B. da Costa, \textit{Senior Member}, \textit{IEEE}, Petar Popovski, \textit{Fellow}, \textit{IEEE}, and Constantinos B. Papadias, \textit{Fellow}, \textit{IEEE}\vspace{-8mm}
}

\maketitle

\begin{abstract}
	Rate-splitting multiple access (RSMA) has recently appeared as a powerful technique for improving the downlink performance of multiple-input multiple-output (MIMO) systems. By flexibly managing interference, RSMA can deliver high spectral and energy efficiency, as well as robustness to imperfect channel state information (CSI). In another development, an intelligent reflecting surface (IRS) has emerged as a method to control the wireless environment through software-configurable, near-passive, sub-wavelength reflecting elements. This article presents the potential of synergy between IRS and RSMA. Three important improvements achievable by IRS-RSMA schemes are identified, supported by insightful numerical examples, and mapped to beyond-5G use cases, along with future research directions.
\end{abstract}

\section{Introduction}
A multiple-input multiple-output (MIMO) system can implement spatial division multiple access (SDMA) to communicate with multiple spatially separated users simultaneously and at the same frequency. This can improve system rate, scalability, reliability and latency, making MIMO an indispensable technology for fifth-generation (5G) communication systems. Nevertheless, by relying solely on SDMA, severe inter-user interference can be experienced when users have overlapping spatial directions or are located close to one another in overload scenarios. To tackle this issue, strategies exploiting different domains have been combined in MIMO systems, from conventional orthogonal multiple access (OMA) techniques, such as time-division multiple access (TDMA) and orthogonal frequency-division multiple access (OFDMA), to non-orthogonal multiple access (NOMA) techniques, such as power-domain and code-domain NOMA. Under ideal conditions, all these techniques efficiently mitigate inter-user interference. However, they are underpinned by an assumption of perfect channel state information (CSI), which is difficult to achieve in real-world deployments. In practice, CSI inaccuracies can diminish the data rates of MIMO, MIMO-OMA, and MIMO-NOMA schemes \cite{RSMAbc21,RSMAbc18}.

Rate-Splitting Multiple Access (RSMA) addresses the drawbacks of OMA and NOMA under imperfect CSI \cite{RSMAbc18}. Unlike SDMA and OMA techniques, which treat residual multi-user interference as noise, or NOMA, which by relying on successive interference cancellation (SIC), fully decodes the interference, the innovative RSMA technique combines the two approaches and flexibly treats one fraction of the interference as noise and addresses the other fraction through SIC. The RSMA technique unlocks a flexible interference management framework that can deliver high spectral and energy efficiencies, optimality in terms of degrees of freedom (DoF), and robustness to imperfect CSI \cite{RSMAbc21}.

Even though RSMA has numerous benefits, establishing reliable communication links through fast-varying wireless channels continues to be a challenge. As an attempt to overcome channel issues, a promising technology called an intelligent reflecting surface (IRS) has been proposed \cite{IRSmre19}. An IRS can be seen as a cluster of controllable scatterers, where each scatterer, i.e., a reflecting element, can be configured independently to generate distinct amplitude and phase responses. Collectively, the reflecting elements of an IRS are able to manipulate and reflect impinging electromagnetic waves with an optimized radiation pattern, creating a large number of possibilities for tuning the propagation medium. Conventionally, these reflections are performed without active amplification. As a result, an IRS does not require amplifiers or other components of conventional radio-frequency (RF) chains, which give IRSs the potential to enable ubiquitous connectivity in beyond-5G at low energy costs. Furthermore, several other advantages have been reported, including advanced control of users' channel gains, extended coverage range and improved fairness \cite{IRSas20}. Due to these features, several works investigating the application of IRS to diverse systems have appeared recently, of which some have focused on the IRS-RSMA topic \cite{IRSRSab21}.

Nevertheless, few technical contributions have been presented and the full possibilities of the combination of IRS and RSMA remain to be investigated. This major literature lacuna motivates the work reported in this paper. Specifically, we perform an in-depth investigation of the possible benefits that combined RSMA and IRS can provide. On the one hand, we show that IRSs can enable a more flexible precoding design and make RSMA resilient to imperfect SIC. On the other hand, we show that RSMA can contribute to robust IRS optimizations even under imperfect CSI. We demonstrate through simulations that these features unleash performance gains unreachable with other multiple access (MA) techniques. We also present future use case scenarios enabled by IRS-RSMA in beyond-5G networks. The paper concludes by considering existing challenges and promising research directions.

\section{An overview on RSMA and IRS Technology}

\subsection{Introduction to the RSMA Technique}
The work in \cite{Cover72} can be seen as one of the first studies to demonstrate that superior rate regions are achievable through rate-splitting (RS) strategies in broadcast channels of single-antenna systems. The goal of achieving new rate regions motivated subsequent works investigating further RS approaches. For instance, the RS strategies presented in \cite{Car78} and \cite{Han81} indicated that if users are being served through interfering channels, it can be beneficial to convey part of the information in a shared common stream and decode part of the inter-user interference. This feature was recently found useful for improving the performance of modern MIMO systems.

Practical MIMO-OMA systems conventionally rely on linear precoding, such as zero-forcing (ZF) precoding, to mitigate inter-user interference. The adoption of these precoding techniques is mainly motivated by their low computational complexity and their optimality in terms of DoF under perfect CSI. In practice, however, the CSI estimate is inevitably imperfect, which makes linear precoding unable to cancel the interference completely and ultimately results in residual noise at the receivers. Such an issue can reduce the system's DoF and limit the achievable data rates. As an attempt to alleviate this limitation, researchers have recently exploited the RS concepts proposed in \cite{Cover72, Car78, Han81} to develop a new robust MA technique for MIMO systems, called RSMA. The technique can be seen as an implementation of the ``{\it divide-and-conquer}'' concept where part of the inter-user interference is addressed at the base station (BS), e.g., through ZF precoding, and part by the users, through SIC. Consequently, it becomes possible to manage how much interference (not canceled by precoding due to imperfect CSI) is treated as noise and how much is decoded. This flexibility makes the technique powerful even in scenarios with inaccurate CSI.

In its simplest form, RSMA splits the data messages of different users into two parts. One part of each message is encoded into a common symbol and the remaining part into private symbols. The common symbol is multiplied by a common precoder (intended for all users), and the private symbols are multiplied by private precoders (each one designed for a particular user). The obtained streams are superimposed in the power domain and then transmitted towards the users. At the receivers' side, all users first decode the common stream, while treating the private streams as noise, and perform SIC to subtract the retrieved messages from the superimposed stream. After SIC, the data in the private streams is finally decoded, ideally, interference-free. Note that unlike NOMA, in which the number of SIC layers increases with the increase of users, RSMA requires that all users (independently of the number) execute SIC only once. Due to this feature, the technique is commonly called single-layer RSMA. Recent results have demonstrated that single-layer RSMA can outperform all conventional OMA, NOMA, and SDMA counterparts \cite{RSMAbc21}. Advanced techniques with multiple common streams and multiple layers of SIC have been also proposed (called generalized RSMA \cite{RSMAbc18}, for instance). However, more complicated schemes are beyond the scope of this paper.

\subsection{Basics of the IRS Technology}
An IRS consists of a thin two-dimensional structure that comprises multiple reflecting elements with adjustable electromagnetic properties. The reflecting elements, made of passive conductive materials, are tuned by a low-power control layer, which can be implemented through diverse technologies. Existing designs propose the use of PIN diodes, varactors, graphene, and liquid crystal-based solutions \cite{IRSmre19}. Moreover, the reflecting elements usually have dimensions much smaller than the carrier wavelength \cite{Dajer20}. Comprising such tiny components enables IRSs to steer signals ideally in any direction and achieve various goals, such as to maximize signal-to-noise ratio (SNR), assist interference cancellation, or operate in absorption mode for security purposes. 

The fast-fading channels of an IRS-assisted system can be represented by the addition of a matrix modeling the direct link between the BS and users and a channel matrix corresponding to the reflected link via the IRS, which, in turn, is usually represented by the dyadic channel model \cite{Liang19}. The dyadic model is a multiplicative channel representation containing three matrices: a matrix for the link between the BS and IRS, a matrix for the link between the IRS and users, and a diagonal matrix with complex-valued elements that model the induced reflections, with amplitudes and phases limited to $[0, 1]$ and $[0, 2\pi]$, respectively. This model is used in the simulation examples given in this article.

\section{IRS-RSMA: Potential Improvements}\label{secpi}
\begin{figure}[t]
	\centering
	\includegraphics[width=1\linewidth]{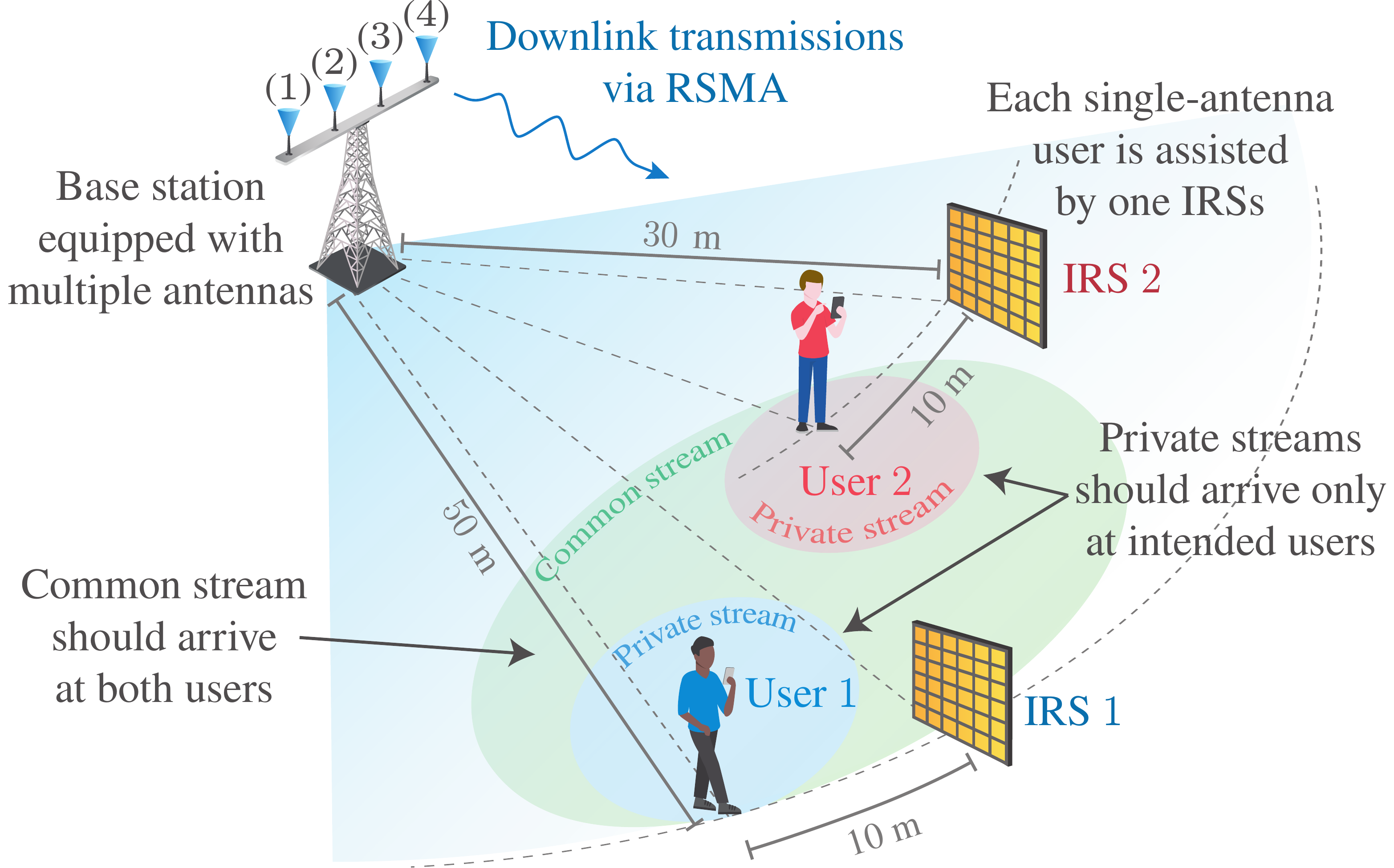}
	\caption{IRS-RSMA setup. Single-antenna users are assisted by IRSs. }\label{f6}
\end{figure}

In this section, we discuss three potential performance improvements that the combined use of IRSs and RSMA offers. Each improvement is supported by representative simulation examples, where we compare the downlink performance of MIMO systems in combination with different MA solutions, including TDMA, NOMA, and RSMA. In all implemented systems, we consider a narrow-band block-fading channel model, where the channel coefficients remain constant during a time slot but change independently over different time slots, such that time dispersion and fading correlation are not modeled. For illustration purposes, in the RSMA schemes, random precoders are employed for broadcasting the common messages, and ZF precoding is used for transmitting the private ones. In the baseline schemes, matched filter (MF) precoding is adopted in the NOMA systems and ZF precoding in the TDMA counterpart. We consider a scenario with two single-antenna users, where user $1$ is located at $50$~m and user $2$ at $30$~m from a BS equipped with $4$ antennas, as shown in Fig. \ref{f6}. Each user is assisted by one IRS containing $50$ reflecting elements, with each IRS located $10$~m apart from its connected user. For simplicity, the distances from the IRSs to the BS are the same as those from the connected users to the BS, and the path-loss exponent is set to $2.5$ in all links. Moreover, the precoders are normalized to unity and fixed power allocation policies are employed in all systems. The power coefficients for users $1$ and $2$ in the NOMA schemes are configured as $\alpha_1 = 7/8$ and $\alpha_2 = 1/8$, whereas in TDMA all available power is allocated in each time-slot to the scheduled user, i.e., $\alpha_1 = \alpha_2 = 1$. In turn, in the RSMA systems, the power coefficients for the private messages are set as $\alpha_1 = \alpha_2 = (1 - \alpha_c)/2$, where $\alpha_c$ represents the coefficient for the common message. It is noteworthy that, even though we consider a narrow-band channel model in the simulations, the gains presented in this section should also be applicable to other system setups. We recognize, though, that new insights could be achieved with different models, such as frequency-selective channels. However, this possibility is left for future work where an in-depth investigation can be carried out.

\subsection{Enhanced Rate of the Common Message}
The concept introduced by RSMA of conveying part of the information through a common stream brings a novel DoF for configuring IRSs, which makes this synergy unique and distinct from what is achievable with classic MA solutions. Specifically, IRSs can assist the design of efficient precoders for broadcasting common messages, which is known to be a challenge in RSMA. Several approaches have been proposed for addressing this issue in conventional RSMA systems \cite{RSMAbc18}. However, most strategies usually favor some users more than others or result in complex optimization problems. Moreover, constructing a single precoder capable of meeting the rate requirement for the common message for all users may be an infeasible task. On the other hand, by assisting common stream transmissions with IRSs, it could be possible to deliver strong signal beams to users even when employing simple precoders at the BS. 

\begin{figure}[t]
	\centering
	\includegraphics[width=1\linewidth]{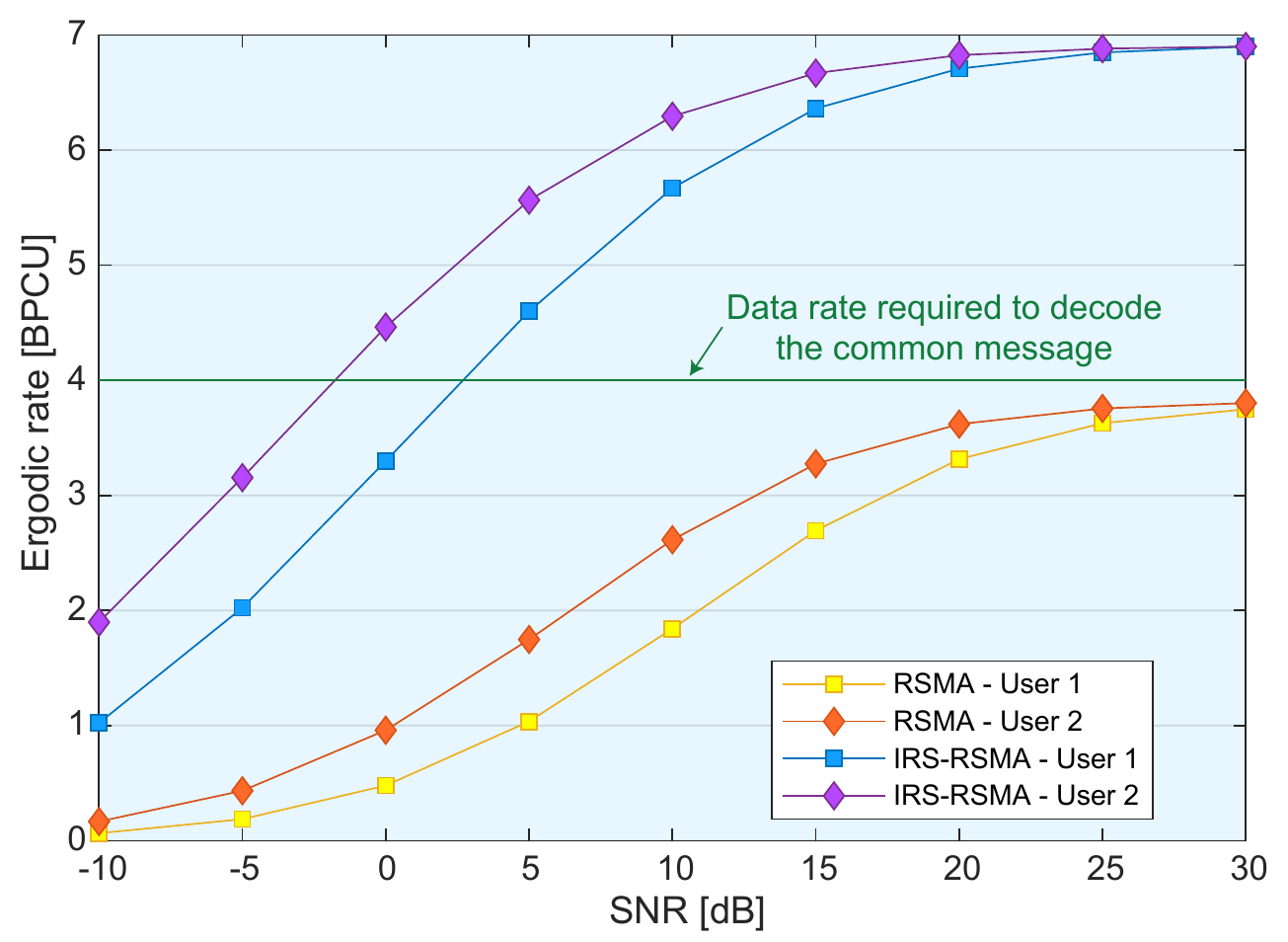}
	\caption{Ergodic rates vs. SNR for the common message in RSMA schemes ($\alpha_c = 0.9, \alpha_1 = \alpha_2 = 0.05$).}\label{f4}
\end{figure}

The above gain is illustrated in Fig. \ref{f4}, where the ergodic rates for the common message achieved with RSMA and IRS-RSMA are presented. The IRS of each user is optimized to match (to add constructively) the channel gains of the common stream achieved in the reflected link with the gains observed in the direct link. As can be seen, due to the interference generated by the private streams, the data rates of the common message become limited in the high-SNR regime in both RSMA and IRS-RSMA schemes. Note, however, that only the two users served via conventional RSMA are not able to meet the minimum rate required to decode the common message. In contrast, when the IRSs are optimized to boost the common message, the users in the IRS-RSMA system become able to achieve a rate higher than the requirement of $4$~bpcu after $5$~dB, reaching almost $7$~bpcu at $30$~dB. This higher rate represents an impressive improvement of more than $3$~bpcu when compared with the rate observed in the RSMA counterpart for the same SNR value. 

\subsection{Robustness to Imperfect SIC}
One key feature shared by NOMA and RSMA is that both techniques rely on SIC to decode part of the transmitted messages. In ideal conditions, it is possible to decode the messages perfectly through SIC without any errors. In practice, however, as a result of hardware imperfections and other impairments, even if the CSI can be perfectly estimated, decoding errors may still occur during the SIC process. It has been demonstrated in the literature \cite{ASIS20} that imperfect SIC can severely harm the performance of NOMA systems and make them less spectrally efficient than conventional OMA schemes. An in-depth investigation of the impacts of imperfect SIC on RSMA schemes is still missing in the literature. Nevertheless, it is evident that despite the benefits of RSMA, its performance can deteriorate as a result of SIC errors. The deployment of IRSs could be highly beneficial to alleviate this issue in RSMA. Specifically, by assisting the RSMA scheme with IRSs and properly splitting the data symbols between common and private messages, high performance can be achieved even under the constraint of imperfect SIC.

\begin{figure}[t]
	\centering
	\includegraphics[width=1\linewidth]{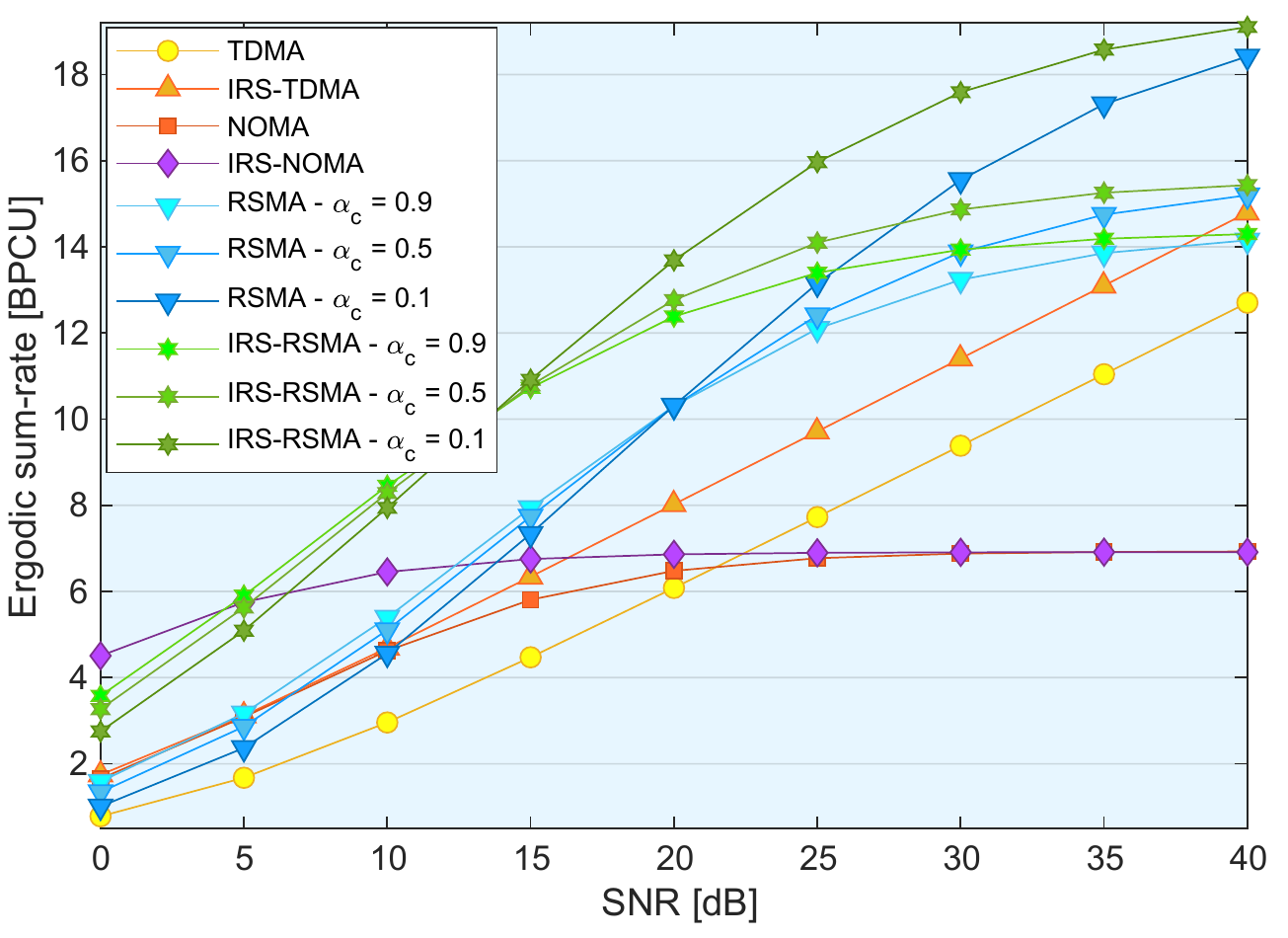}
	\caption{Impact of imperfect SIC on the ergodic sum-rates of various MA systems (SIC error factor $= 0.01$).}\label{f2}
\end{figure}

The robustness of IRS-RSMA to residual SIC errors is illustrated in Fig. \ref{f2}, where the IRSs are optimized to boost the users' channel gains and mitigate interference through a constrained least-squares approach. As in \cite[Sec. III]{ASIS20}, a deterministic error factor is used to model the residual interference left by imperfect SIC. It can be seen that both the NOMA and the IRS-NOMA system are strongly impacted by residual SIC errors, with sum-rates limited to only $6.9$~bpcu in the high-SNR regime, which is inferior to that achieved by the TDMA counterpart after $20$~dB. In contrast, despite the presence of SIC errors, the sum-rates of the RSMA schemes can outperform the TDMA counterpart in the entire considered SNR range. For SNR values below $10$~dB the RSMA system can, however, still be outperformed by its NOMA and IRS-TDMA counterparts. On the other hand, the IRS-RSMA system can overcome this issue and boost the sum-rate in almost all the SNR range. For instance, when the SNR is $25$~dB and the power allocated to the common message is $\alpha_c = 0.5$, the IRS-RSMA scheme can achieve a sum-rate of $14$~bpcu, which is superior to that achieved by the RSMA counterpart for all values of $\alpha_c$, and more than $6$~bpcu higher than the sum-rate achieved by the TDMA and NOMA-based schemes. 

\subsection{Robustness to Imperfect CSI}
As IRSs comprise only nearly passive components, accurately estimating the channels of the cascade-reflected link, i.e., the channels between the BS and the IRS, and the IRS and users, has been one of the main challenges of IRS-assisted systems, and different strategies have been proposed. Most common approaches try to estimate the full concatenated channel entirely at the BS through uplink training techniques, and some other approaches try to simplify the estimation process by installing scattered active sensors in the IRS \cite{Dajer20}. Nevertheless, independently of the IRS hardware or estimation strategy, perfectly obtaining the global CSI remains a complicated task. Consequently, in practice, the optimization of the IRSs is usually performed based on imperfect CSI, which might result in sub-optimal performance. Fortunately, unlike other conventional MA techniques, RSMA has the advantage of being robust in scenarios with inaccurate channel estimation. Therefore, RSMA is a good fit for realistic IRS-assisted systems with imperfect CSI.

\begin{figure}[t]
	\centering
	\includegraphics[width=1\linewidth]{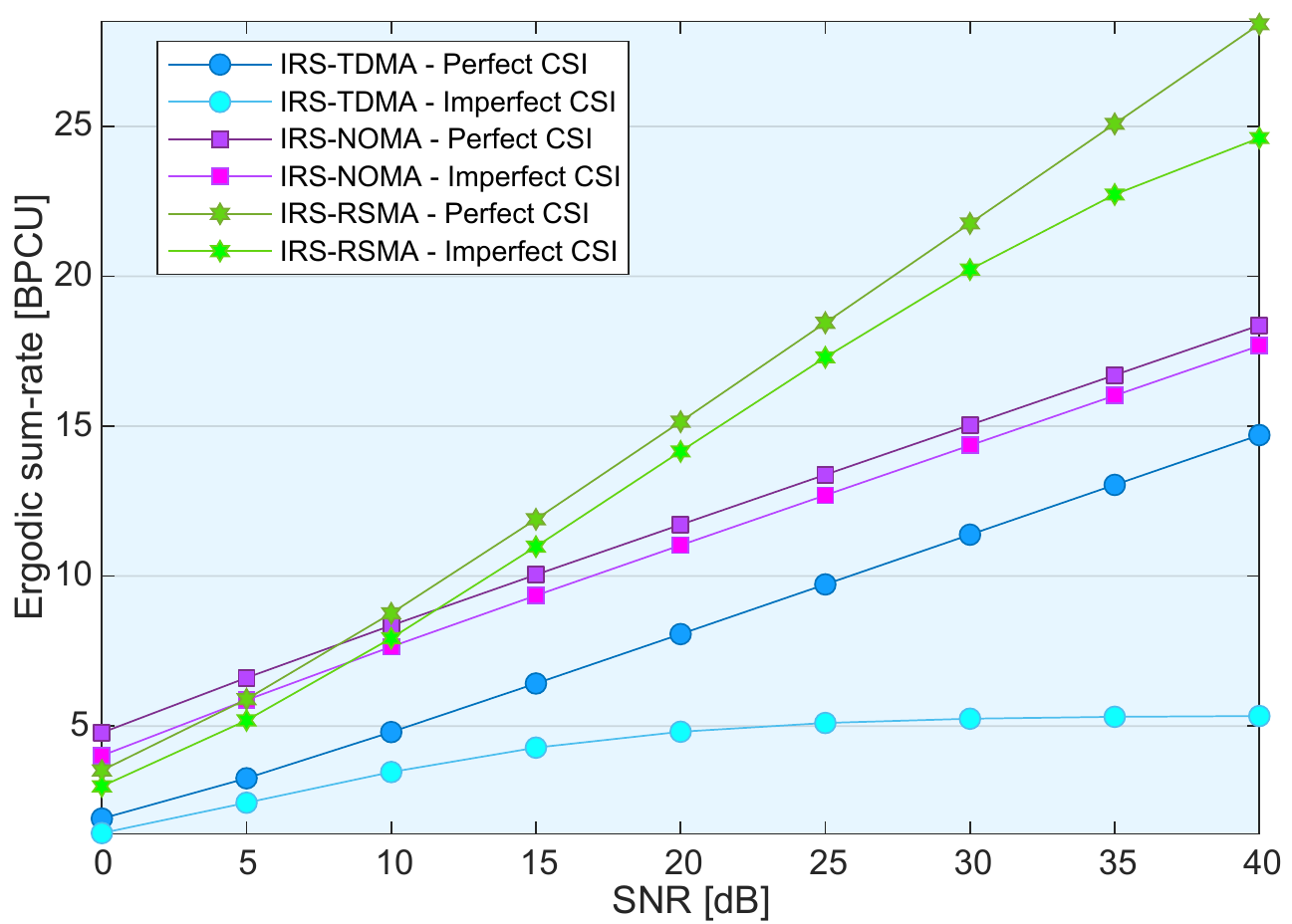}
	\caption{Ergodic sum-rate curves for various MA systems under imperfect CSI (channel error variance $= 0.5$, $\alpha_c = 0.9, \alpha_1 = \alpha_2 = 0.05$).}\label{f3}
\end{figure}

Fig. \ref{f3} plots the ergodic sum-rates of various IRS-assisted systems under perfect and imperfect CSI. In this figure, the reflecting elements of the IRSs are also optimized to boost the users' channel gains and mitigate interference. Note that because ZF precoding is unable to cancel the inter-user interference completely when the CSI is imperfect, the IRS-TDMA is the most impacted scheme, with its sum-rate saturating slightly above $5$~bpcu at high SNR, which is almost $10$~bpcu lower than the IRS-TDMA scheme can reach at $40$~dB under perfect CSI. On the other hand, the IRS-RSMA system can achieve high sum-rate levels with perfect and imperfect CSI, outperforming all the baseline schemes. For instance, when the SNR is $30$~dB, the IRS-RSMA scheme with perfect CSI can obtain a sum-rate of $21.7$~bpcu, while under imperfect CSI it can still achieve an impressive $20.2$~bpcu, which is approximately $6$~bpcu higher than that achieved by the IRS-NOMA counterpart with imperfect CSI, and $15$~bpcu higher than that of the IRS-TDMA counterpart with imperfect CSI. These results confirm that by combining RSMA and IRS technology, it becomes possible to deploy robust communication systems even when the channel estimation is poor.

\begin{figure*}[t]
	\centering
	\includegraphics[width=1\linewidth]{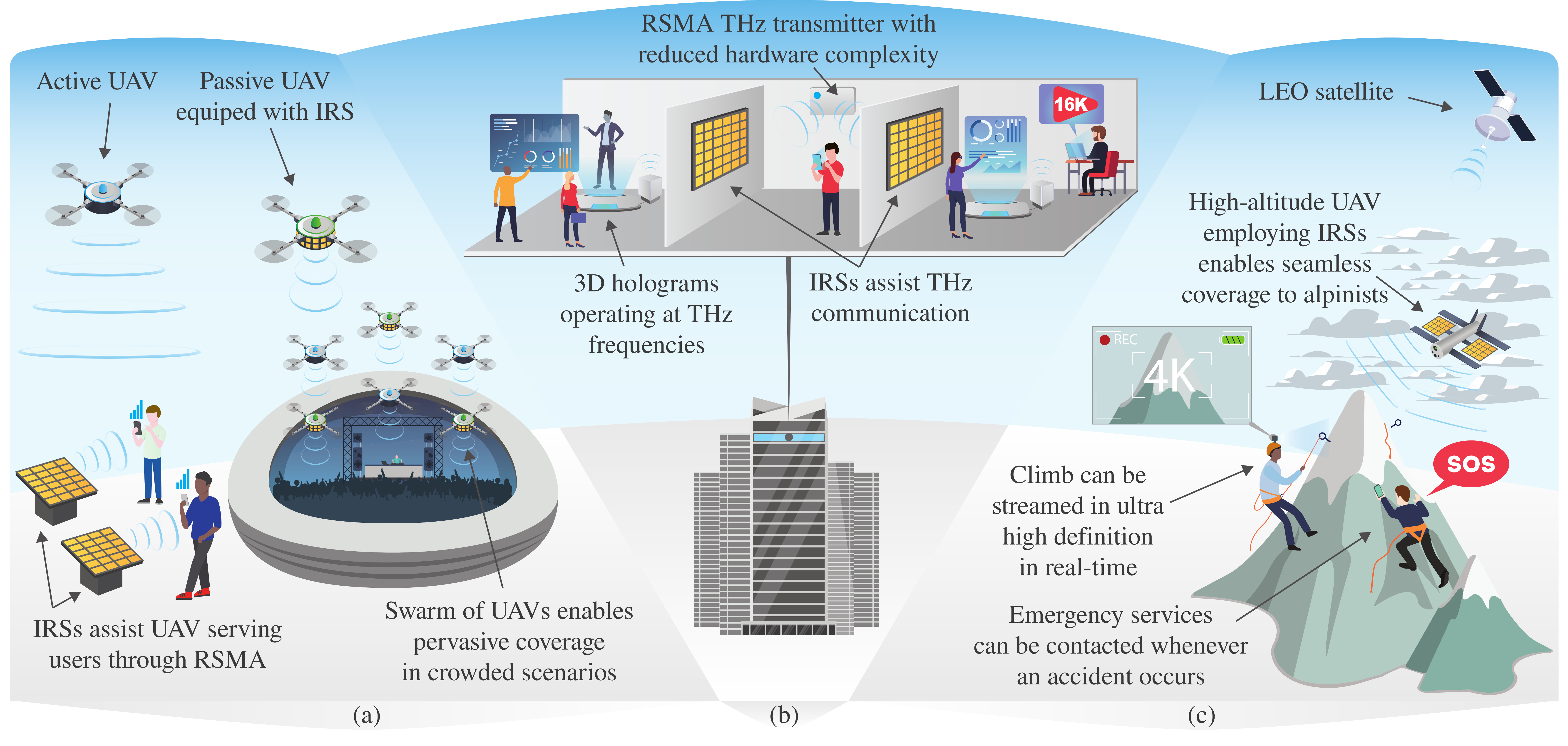}
	\caption{Potential use case scenarios enabled by IRS-RSMA in future wireless networks beyond 5G.
	}\label{f5}
\end{figure*}

\section{Potential Use Cases in Beyond-5G}
In this section, we present use case scenarios that can be enabled with the aid of IRS-RSMA schemes.

\subsection{CSI-Robust UAV Networks}
Unmanned aerial vehicles (UAVs) will play an important role in beyond-5G \cite{Jaafar20}. UAVs have been envisioned as executing diverse tasks, ranging from goods deliveries, surveillance and military applications, to working as BSs, where UAVs can provide flexible and dynamic coverage. For instance, with the help of UAVs, it will be possible to temporarily support high-performance connectivity in hyper-crowded environments, such as in stadiums and at festivals, or extend signal coverage to remote rural areas and during natural disasters. The deployment of swarms of UAVs is another promising application for future communication systems. In a swarm, by sensing the environment, a large number of UAVs can intercommunicate, reorganize and adapt autonomously in the air, allowing them to execute even the most complex tasks.

Nevertheless, there are unsolved issues that need to be tackled before UAV networks become everyday reality. IRS-RSMA schemes can efficiently address some of these challenges. In particular, channel interference and the overhead generated by multiple UAVs and their fast 3D motion make channel estimation a complicated process, which can potentially result in imperfect CSI. As noted earlier, in addition to being effective at managing interference, IRS-RSMA schemes are robust to imperfect CSI, which makes them very suitable for employment in UAV networks. For instance, RSMA could enable multiple UAVs to communicate efficiently with multiple ground users, whereas IRSs pointed to the sky could guarantee a strong communication link between users and UAVs. One can also envision a CSI-robust swarm of UAVs equipped with IRSs, with some UAVs active and others passive, interacting with each other via RSMA. As illustrated in Fig. \ref{f5}a, in both examples, the UAVs would be able to deliver high data rates seamlessly with low impact from a possibly degraded CSI.

\subsection{Enhanced High-Frequency Communication}
5G has expanded its operating bandwidth by adopting higher frequency bands above $6$~GHz, known as the millimeter-wave (mmWave) spectrum. While 5G is expected to operate at frequencies not higher than $100$~GHz, 6G and beyond generations are foreseen to go further and operate from the upper band of the mmWave spectrum ($100$~GHz to $300$~GHz), up to the terahertz (THz) spectrum (up to $3$~THz) \cite{Rikkinen20}. However, due to the high signal attenuation and absorption, operating at such high frequencies becomes very challenging. The coverage range of mmWave and THz communication can be improved by deploying a large number of transmit antennas (hundreds or thousands). Nevertheless, the large number of RF chains required in these large-scale arrays may lead to complicated precoding design and excessive feedback overhead.

On the other hand, by deploying IRSs to induce sharp beams directed towards the users, it becomes possible to mitigate signal attenuation and extend signal coverage. In turn, by leveraging the RSMA technique, due to its robustness to imperfect CSI, the required CSI feedback overhead can be reduced without significantly deteriorating the system performance. These capabilities suggest that IRS-RSMA schemes can help to reduce the required number of active antennas and RF chains at the BS and still deliver a good communication performance with the benefits of reducing the BS hardware complexity. Moreover, since IRSs comprise only low-power components, the energy required for optimizing the reflecting elements should be less than the energy savings achieved with a fewer number of power-hungry RF chains at the BS. Therefore, as another benefit, IRS-RSMA could also reduce the overall energy consumption of mmWave and THz systems. As illustrated in Fig. \ref{f5}b, a BS of an indoor THz IRS-RSMA system could support real-time 3D holographic meetings in multiple rooms of a commercial building and, at the same time, stream $16$K resolution video to users' laptops and advanced smartphones.

\subsection{Seamless Satellite Communication}
Enabling seamless connectivity across the entire Earth's surface, from high altitude mountain ranges to the middle of the oceans, by relying on terrestrial and UAV networks alone may be unrealizable or too costly. On the other hand, low earth orbit (LEO) satellite networks can cover vast geographical areas and potentially deliver high data rates to the most remote and inhospitable regions of the globe \cite{You20,Matthiesen21}. Yet, LEO networks come with some drawbacks. Specifically, due to the atmospheric gases, rain, and cloud coverage, the signals transmitted and received by satellites can suffer strong attenuation. Moreover, due to long-distance signal travel, the CSI available in satellites can become outdated.

IRS-RSMA strategies are also suited to cope with these satellite-related communication issues. First, by employing the RSMA technique, LEO satellites become able to deliver high data rates to multiple users even when the CSI is outdated. Second, by deploying IRSs to boost the signal transmissions, it is possible to mitigate the effects of atmospheric phenomena. For example, high-altitude UAVs equipped with IRSs could be deployed near cloud cover to assist a LEO network to serve multiple alpinists via RSMA in remote mountainous regions. This IRS-assisted RSMA-LEO network would ensure high data rates and reduce the probability of alpinists finding themselves out of coverage. Such capabilities would enable alpinists to stream their climb in real-time even under adverse weather conditions. These features could also save lives following accidents. Fig. \ref{f5}c illustrates this extreme scenario.

\section{Challenges, Future Directions and Concluding Remarks}
We have demonstrated that the use of IRSs and RSMA can bring mutual benefits, in that IRSs can help to address issues related to precoding design and imperfect SIC in RSMA, and, reciprocally, RSMA can bring robustness to the imperfect CSI that is unavoidable in IRS-assisted communications. However, both RSMA and the IRS technology are still in their infancy, and there are still several open problems that need to be solved before practical deployment can happen. Specifically, more extensive studies need to be carried out to determine the best optimization strategies, the most appropriate IRS architecture, and associated tradeoffs. Frequency-selective fading in IRS-RSMA still needs to be better investigated. Furthermore, due to the more complex encoding, IRS-RSMA schemes require new signaling strategies to coordinate users, IRSs and the BS, and extensive tests need to be performed in real-world test-beds to confirm the claimed gains.

The combination of IRS and RSMA is a research area with exciting possibilities for future work. IRS-RSMA contributions investigating the application of different modulation techniques, aerial networks, and THz communications, spanning from performance analysis and resource allocation to channel estimation strategies, are interesting and exciting future directions. All in all, the full potential of the interplay of IRSs and RSMA is yet to be realized.

\section*{Biographies}\vspace{-0mm}
	\vskip -0\baselineskip plus -.5fil
	\begin{IEEEbiographynophoto}{Arthur Sousa de Sena}[M] (arthur.sena@lut.fi) is with Lappeenranta-Lahti University of Technology, Finland.
	\end{IEEEbiographynophoto}
	
	\vskip -.5\baselineskip plus -.5fil

	\begin{IEEEbiographynophoto}{Pedro H. J. Nardelli}[SM] (pedro.nardelli@lut.fi) is with Lappeenranta-Lahti University of Technology, Finland.
	\end{IEEEbiographynophoto}
	
	\vskip -.5\baselineskip plus -.5fil

	\begin{IEEEbiographynophoto}{Daniel Benevides da Costa}[SM] (danielbcosta@ieee.org) is with the Technology Innovation Institute (TII), Abu Dhabi, United Arab Emirates.
	\end{IEEEbiographynophoto}
	
	\vskip -.5\baselineskip plus -.5fil
	
	\begin{IEEEbiographynophoto}{Petar Popovski}[F] (petarp@es.aau.dk) is with Aalborg University, Denmark.
	\end{IEEEbiographynophoto}
	
	\vskip -.5\baselineskip plus -.5fil

	\begin{IEEEbiographynophoto}{Constantinos B. Papadias}[F] (cpapadias@acg.edu) is with the American College of Greece, Greece.
	\end{IEEEbiographynophoto}

\end{document}